\documentclass[
reprint,
groupedaddress,
twocolumn,
 amsmath,amssymb,
 aps,
prb,
floatfix
]{revtex4-2}

\usepackage{textcomp}
\usepackage{graphicx}
\usepackage{dcolumn}
\usepackage{bm}
\usepackage{float}
\usepackage[colorlinks, allcolors=blue]{hyperref}
\usepackage{todonotes}
\usepackage{tikz}

\usetikzlibrary{decorations.pathreplacing}
\usetikzlibrary{arrows.meta}
\tikzset{x=1in,y=1in,z=1in}

\newcommand*{\figref}[2][]{\ref{#2}\hyperref[{#2}]{#1}}

\usepackage[normalem]{ulem} 

\begin{document}


\title{Enhanced repulsive Casimir forces between gold and thin magnetodielectric plates}

\author{C. Shelden}
 \affiliation{Department of Electrical and Computer Engineering, University of California, Davis, CA 95616, USA}

\author{B. Spreng}
\affiliation{Department of Electrical and Computer Engineering, University of California, Davis, CA 95616, USA}

\author{J. N. Munday}
\email{jnmunday@ucdavis.edu}
\homepage{https://mundaylab.com}
\affiliation{Department of Electrical and Computer Engineering, University of California, Davis, CA 95616, USA}

\date{June 13, 2023}

\begin{abstract}

We calculate repulsive Casimir forces between metallic and magnetic plates and quantitatively probe the magnetic plate's properties as tuning knobs for the repulsion. Namely, the plate's thickness and its low-frequency permittivity and permeability. For a thin magnetic plate ($\leq 10\,\text{nm}$), we find that repulsion can exist as long as $\mu(0) \geq \epsilon(0)$. We also explore the effect of temperature on the repulsion and transition distance between attractive and repulsive interactions. We show how the parameters can be tuned to allow repulsion at sub-micron separation regimes, making it potentially accessible to known high-resolution measurement techniques using magnetic van der Waals materials.

\end{abstract}

\maketitle


\section{Introduction} \label{sec:Introduction}

In 1948, Hendrick Casimir theorized that two perfect conducting plates positioned parallel to each other in a vacuum would experience an attractive force arising from the zero-point energy of the vacuum \cite{casimir_attraction_1948}, since referred to as the \textquotesingle Casimir force\textquotesingle. Evgeny Lifshitz generalized this work to account for the material's real optical response \cite{lifshitz_theory_1956}, which was further expanded by Dzyaloshinskii, Lifshitz, and Pitaevskii to include the thermal fluctuations present at finite temperatures \cite{dzyaloshinskii_general_1961}.

Experimentally generating a repulsive Casimir force between two plates separated by less than a micron in vacuum has eluded the community to date. There have been a few theoretical approaches to potentially overcoming this limitation, which include leveraging nonequilibrium states \cite{antezza_casimir-lifshitz_2006, antezza_casimir-lifshitz_2008, kruger_non-equilibrium_2011, golyk_casimir_2012, chen_nonequilibrium_2016}, exotic geometries and materials \cite{levin_casimir_2010, jiang_chiral_2019, zhao_repulsive_2009, butcher_casimirpolder_2012}, and magnetic properties \cite{kenneth_repulsive_2002, henkel_casimir_2005,  tomas_casimir_2005, rosa_casimir-lifshitz_2008, rosa_casimir_2008, yannopapas_first-principles_2009, geyer_thermal_2010, inui_thickness_2011, zhao_repulsive_2011, inui_quantum_2012, brevik_casimir_2018}. This article focuses on the conditions by which magnetic materials may be used to create repulsive forces and the extent to which currently available materials could be used.

T. H. Boyer derived the force between a perfect conducting plate and an infinitely permeable plate \cite{boyer_van_1974}, which resulted in a purely repulsive force with a magnitude slightly below that which Casimir calculated for the attractive case. Since then, the question has arisen, `how ideal must the materials be to produce repulsion?' In real systems, the interaction is not purely repulsive, but rather it can be attractive at some separations and repulsive at others \cite{tomas_casimir_2005, rosa_casimir-lifshitz_2008, inui_quantum_2012}. This aspect factors into material choices as more ideal materials, especially magnetic materials, can push the repulsive behavior to shorter separations. Initial generalizations investigated nondispersive media \cite{kenneth_repulsive_2002}. Although informative for the first steps of real material implementation, it was noted that dispersion must be included moving forward \cite{iannuzzi_comment_2003}.

With dispersion accounted for, the repulsive component has been shown to be sensitively dependent on the frequency regime over which the magnetodielectric plate's response is mainly magnetic. Further inspection of force behavior, when accounting for frequency dependence \cite{tomas_casimir_2005}, showed similar results to the nondispersive case, although the repulsion generally occurs at a separation several orders of magnitudes below the length-scale corresponding to the wavelength of the magnetic resonance frequency. This result first signaled that for repulsion at sub-micron separations, the magnetic resonance may not necessarily need to occur at visible frequencies.

Despite the results of Ref.~\cite{tomas_casimir_2005}, in many other systems, the Casimir force at short separations is dictated by the visible frequency response of the interacting materials.  
With this in mind, metamaterials have been proposed as a way of producing an artificial magnetic response~\cite{henkel_casimir_2005, tomas_casimir_2005, rosa_casimir-lifshitz_2008, rosa_casimir_2008, yannopapas_first-principles_2009}. This property is useful because naturally-occurring materials do not have a magnetic response at visible frequencies \cite{iannuzzi_comment_2003}.

Reference~\cite{rosa_casimir-lifshitz_2008} probed the connection between the repulsive behavior and the properties of a magnetic metamaterial and found that even a small Drude background in the permittivity of the magnetic metamaterial causes the repulsive component to turn attractive in the $\mu$m separation regime but that repulsion can be enhanced by increasing the magnetic anisotropy of the metamaterial. They also found that the magnetic response in the visible regime is not necessarily required to produce a Casimir repulsion at sub-micron separations, confirming the results in Ref.~\cite{tomas_casimir_2005}. However, the authors eventually concluded that measured repulsion (in the relevant $d = 0.1$--$1 \, \mu$m separation regime) using magnetic metamaterials would be very challenging \cite{rosa_casimir_2008}. 

Revisiting naturally-occurring materials, it has been shown that Casimir repulsion is still  theoretically possible in metallic-magnetodielectric systems if the magnetic plate has a significantly strong response and the metallic plate's permittivity at vanishing frequency is described by the dissipationless plasma model~\cite{geyer_thermal_2010, inui_quantum_2012}. The latter requirement can be understood when considering that the zero-frequency transverse electric mode provides the sole repulsive contribution to the pressure. In contrast, a dissipative Drude model predicts a vanishing pressure contribution for the transverse electric mode at zero-frequency leaving a purely attractive interaction. While it is an experimental fact that the conductivity of metals such as gold remains finite at low frequencies, which is in agreement with the Drude model, several Casimir experiments are well described by the plasma model, excluding the Drude model with high confidence~\cite{mostepanenko_casimir_2021, klimchitskaya_current_2022}. These results suggests that it is likely that a more complex model, rather than a simple Drude or plasma model, is actually needed to describe the dielectric response of metals for real materials. Thus, metallic-magnetodielectric material systems provide an excellent platform to further experimentally distinguish between the two models as the sign of the Casimir force depends crucially on the modeling as described above. It should be noted that repulsion has also been shown to be present when replacing the metallic plate with a superconducting material whose permittivity can be described with Lorentz oscillators; however, this effect is caused by its diamagnetic properties, which is a separate but interesting system in its own right~\cite{inui_diamagnetic_2011}.  

For a finite thickness magnetodielectric plate, its thickness $b$ and permittivity $\epsilon(\omega)$ and permeability $\mu(\omega)$ at zero-frequency have all been shown to affect the Casimir repulsion~\cite{geyer_thermal_2010, inui_thickness_2011, zhao_repulsive_2011, inui_quantum_2012}. The system temperature $T$ has also been shown to enhance repulsion in metamaterial systems through the zero-frequency transverse electric mode~\cite{rosa_casimir_2008}.

Here, we systematically investigate to what extent each of the preceding parameters can be tuned to enhance the repulsive behavior in a metallic-magnetodielectric system using the largest-magnitude repulsion $P_{\text{max}}$ and separation where the interaction switches from attraction to repulsion $d_{\text{T}}$ as quantitative metrics. This investigation is carried out with the overarching goal of identifying combinations of parameter values that theoretically push this system into an experimentally accessible regime and subsequently allow for the measurement of a Casimir repulsion in vacuum at sub-micron separations for the first time. Although Casimir force calculations using ultra-thin yttrium iron garnet (YIG) as the magnetodielectric material have shown repulsion in the sub-micron separation regime \cite{inui_thickness_2011, inui_quantum_2012}, it is unclear if the bulk properties of YIG will hold when thinning the material down to just a few nanometers. We therefore re-evaluate the parameter space and use our calculations to identify a potential class of alternative materials that could enable repulsion.

  We compare our calculations with the force-modulated gradient measurement technique, which is generally performed using an atomic force microscope (AFM) \cite{de_man_casimir_2010, garrett_measurement_2018, garrett_sensitivity_2019}. The theoretical limits to the minimum force and maximum separations over which this technique can work have been identified in Ref.~\cite{garrett_sensitivity_2019}. We also only consider Au as the material for the metallic plate, as it is the most common surface used in these measurements, and other noble metals generally give similar results.

In addition to the quantitative tuning of already-established connections between the stated system parameters and repulsive behavior, we show a nontrivial interplay between the plate thickness and $\mu(0)$, as well as linearization of the relationship between $\mu(0)$ and $\epsilon(0)$ required to generate repulsion ($\mu(0) \geq \epsilon(0)$), previously seen to be nonlinear for infinitely-thick plates \cite{geyer_thermal_2010}.

The content of this article is partitioned into the following sections: In Sec.~\ref{sec:LifshitzFormalism}, a generic Au-magnetodielectric system, the underlying theory, and key features in the force behavior are discussed; in Sec.~\ref{sec:RelaxedIdeality}, we probe the effect of non-ideality of the magnetodielectric material; in Sec.~\ref{sec:TuningMagnetic}, we look at what relative permeability and thickness regimes are most optimal for repulsion; in Sec.~\ref{sec:ThermalEffects}, the system temperature is considered as an additional tuning knob; Lastly, in Sec.~\ref{sec:Conclusions}, we draw conclusions as to what combination of parameter values appear optimal for the largest repulsion to occur at the smallest separation and identify a class of materials that are excellent candidates for meeting these goals.

\section{Theory and Application} \label{sec:LifshitzFormalism}

\begin{figure}
	\hspace{0em}
	\includegraphics{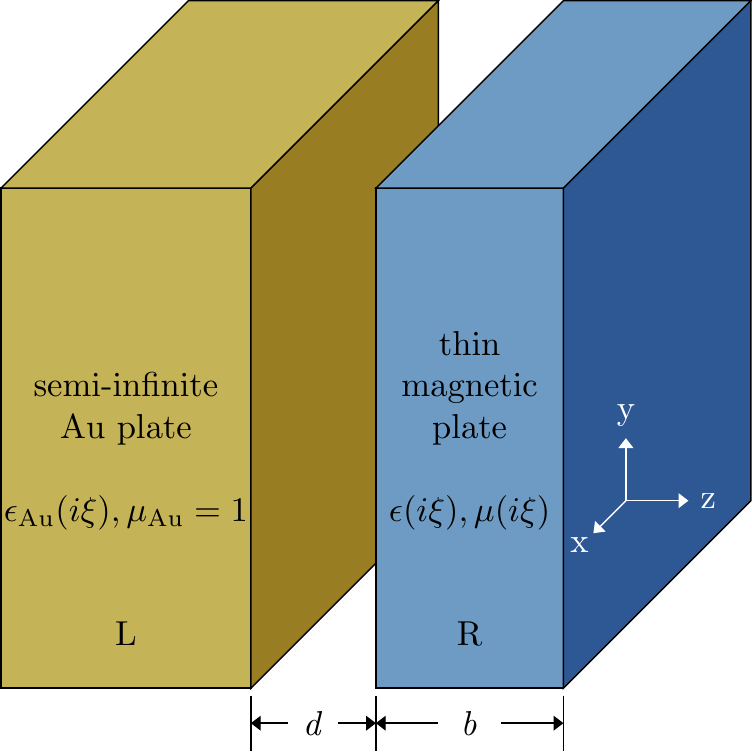}
	\caption{Parallel plates configuration. Left plate is semi-infinitely thick Au; right plate is an arbitrary magnetic material. The material permittivities $\epsilon$ and permeabilities $\mu$ along with the plate-plate separation $d$ and magnetic plate thickness $b$ are noted.}
	\label{fig:plate_configuration_figure}
\end{figure}

We consider a cavity in vacuum formed between a semi-infinite Au plate and a magnetodielectric plate of thickness $b$, positioned parallel to each other and separated by a distance $d$ (Fig.~\ref{fig:plate_configuration_figure}).

\subsection{Lifshitz formalism}

Within Lifshitz's theoretical framework \cite{lifshitz_theory_1956, dzyaloshinskii_general_1961}, the pressure between the plates can be separated into contributions from the transverse electric (TE) and transverse magnetic (TM) modes described by the following equation:
\begin{equation}
	P= P_{\text{TE}} + P_{\text{TM}}\,,
	\label{eqn:pressure}
\end{equation}
where the components of Eq.~(\ref{eqn:pressure}) take the form:
\begin{equation} 
	P_{j}= -\frac{k_{B}T}{\pi} \sum_{n=0}^{\infty}{}^{'} \int^{\infty}_{0} k \, \text{d} k  \,  \frac{r^{\text{L}}_{j} r^{\text{R,eff}}_{j} \; e^{-2 \rho_{\text{m}} d }}{1- r^{\text{L}}_{j}r^{\text{R,eff}}_{j}e^{-2 \rho_{\text{m}} d }} \rho_{\text{m}}\,,
	\label{eqn:Pj}
\end{equation}
and $k_{B}$ is the Boltzmann constant, $T$ is the system temperature, $k$ is the magnitude of the wave-vector projection onto the surface of the plate(s), and $r^{\text{L}}_{j}$ and $r^{\mathrm{R,eff}}_{j}$ are the Fresnel reflection coefficients for polarization $j$ at the left, L (Au), and right, R (magnetic material), plate-surfaces, respectively. 

Taking the finite thickness $b$ of the magnetic plate into account, its effective reflection coefficient is \cite{parsegian_van_2005}:
\begin{equation} \label{eqn:eff_fresnel}
	r^{\text{R,eff}}_{j}=\frac{  r_{j}^{\text{R}}(1 - e^{-2 \rho_{_{\text{R}}} b})}
	{1 - (r_{j}^{\text{R}})^{2} \: e^{-2\rho_{_{\text{R}}} b}}\,.
\end{equation}
The Fresnel reflection coefficients for half-spaces have the following forms:
\begin{equation} \label{eqn:fresnel_coeffs}
	r_{\text{TM}}^{i}=\frac{\rho_{\text{m}}\epsilon_{i}-\rho_{i}}{ \rho_{\text{m}}\epsilon_{i}+\rho_{i}} \; , \;  r_{\text{TE}}^{i}=\frac{\rho_{\text{m}}\mu_{i}-\rho_{i}}{\rho_{\text{m}}\mu_{i}+\rho_{i}}\,,
\end{equation}
where $\epsilon_{i}$ and $\mu_{i}$ are the relative permittivity and permeability of media $i$ (L: Au, m: vacuum, and R: magnetic material), respectively. $\rho_{i}$ is the magnitude of the imaginary component of the wave-vector projected onto the z-axis (normal to the plate surfaces) in media $i$ and has the form:
\begin{equation} 
	\rho_{i}= \sqrt{k^{2}+\frac{\epsilon_{i}\mu_{i} \xi_{n}^{2}}{c^{2}}}\,,
\end{equation}
where $c$ is the speed of light in vacuum. The summation in Eq.~\eqref{eqn:Pj} is taken over the Matsubara frequencies $\xi_{n}=2 \pi n k_{B} T/\hbar$, where $\hbar$ is the reduced Planck's constant and the prime denotes that the zero-frequency term $(n = 0)$ is multiplied by an additional factor of 1/2.

We describe the relative permittivity of the Au plate by the plasma model in the imaginary frequency domain as
\begin{equation}
	\epsilon_{\text{Au}}(i\xi)=1+\frac{\omega_{p}^{2}}{\xi^{2}}\,,
	\label{eqn:plasma}
\end{equation}
using the plasma frequency $\omega_{p} = 9.0 \, \text{eV}$, as in Refs.~\cite{klimchitskaya_casimir_2000, bimonte_beyond-proximity-force-approximation_2018}. Au is nonmagnetic, so the permeability of the Au plate is set to $\mu_{\text{Au}}(i\xi)=1$.

\begin{figure}
	\centering
	\includegraphics[width=0.5\textwidth]{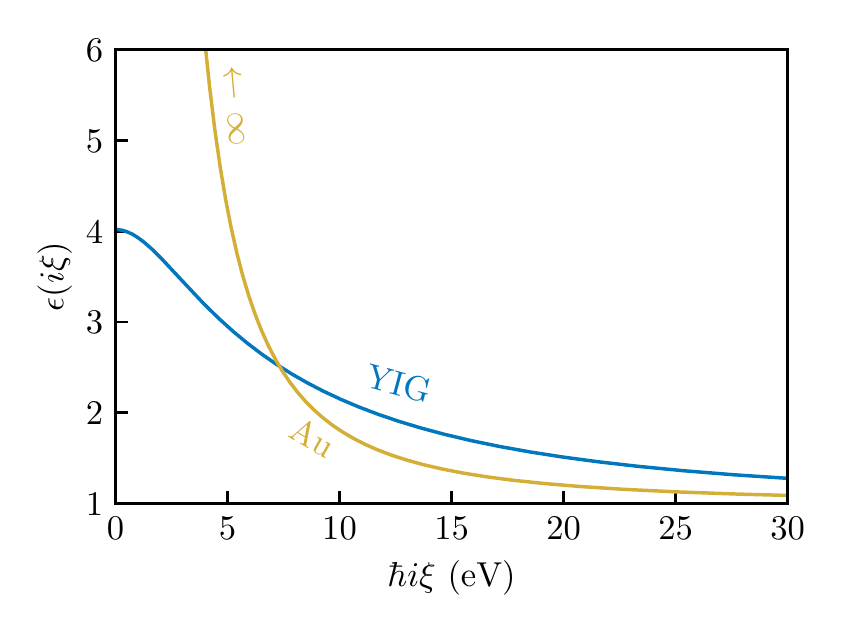}
	\caption{Permittivities for Au and YIG evaluated at imaginary frequencies $i\xi$.}
	\label{fig:permittivities}
\end{figure}

While we will concentrate on the effect of the magnetic material's permeability, its permittivity cannot be ignored (or simply set to 1) for a realistic material. We have chosen to model the permittivity of the magnetic material after that of bulk YIG. The experimental data for the YIG permittivity along with the upper and lower frequency extrapolations in Ref.~\cite{inui_quantum_2012} are used with Kramers-Kronig relation~\eqref{eqn:KKR} to calculate $\epsilon(i\xi)$ for YIG as:
\begin{equation}
	\epsilon(i \xi) = 1+\frac{2}{\pi} \int^{\infty}_{0} \frac{\omega \; \text{Im}[\epsilon(\omega)]}{\omega^{2}+\xi^{2}} \text{d} \omega \,.
	\label{eqn:KKR}
\end{equation}
As seen in Fig.~\ref{fig:permittivities}, the permittivities of Au and YIG are infinite and finite, respectively, at $i\xi = 0$ as expected from a metal and a dielectric. This choice is reasonable but constrains the scope to magnetic materials with a dielectric character (i.e., materials with a finite permittivity at $\xi=0$), which is necessary for a more thorough investigation of the other magnetic material parameters.

Unlike the permittivity, where some materials could be described with a Drude/plasma background and/or resonances in the visible frequency regime, the permeability of naturally-occurring ferromagnetic materials typically only have a response up to the few-GHz frequency regime \cite{ghodgaonkar_free-space_1990}. It is therefore reasonable for system temperatures near $300\, \text{K}$ to treat the magnetic response as quasistatic $\mu(i\xi=0) > 1$ and otherwise $\mu(i\xi > 0)=1$. For a more in-depth discussion on this treatment, please see Ref.~\cite{geyer_thermal_2010}. For most of the work presented here, the value of $\mu(i\xi=0)$ will be left arbitrary rather than pinned at the value for YIG [$\mu(0)=160$], allowing us to explore its dependence on the force.


Special care must be taken for the pressure expressions at $i\xi = 0$ as divergences and indeterminate values emerge in Eq.~\eqref{eqn:Pj} during numerical evaluation. In the limit of $i\xi \rightarrow 0$, Eqs.~\eqref{eqn:Pj} and (\ref{eqn:eff_fresnel}) can be rewritten such that $\rho_{\text{m}}$ and $\rho_{\text{R}}$ are both replaced with $k$ and the Fresnel reflection coefficients~\eqref{eqn:fresnel_coeffs} at the left plate surface take the forms:
\begin{equation}\label{eq:r^L}
	r_{\text{TE},0}^{\text{L}} = \frac{ k -\sqrt{k^{2}+\omega_{p}^{2}/c^{2}}}{k +\sqrt{k^{2}+\omega_{p}^{2}/c^{2}}}
	\; , \; 
	r_{\text{TM},0}^{\text{L}} = 1 \, ,
\end{equation}
and for the right plate surface:
\begin{equation}
	r_{\text{TE},0}^{\text{R}} = \frac{\mu(0)-1}{\mu(0)+1} 
	\; , \; 
	r_{\text{TM},0}^{\text{R}}  = \frac{\epsilon(0)-1}{\epsilon(0)+1} \, .
\end{equation}

\subsection{Typical pressure characteristics}

One of the general requirements for generating a repulsive Casimir force in metallic-magnetodielectric systems is that the permittivity of the electric plate be described by a plasma model at  zero-frequency. We therefore use Au for the metallic plate as it is a good, nonmagnetic metal and is common in experiments \cite{garrett_measurement_2018, garrett_sensitivity_2019}.
 
 \begin{figure}
 	\centering
 	\includegraphics[width=0.5\textwidth]{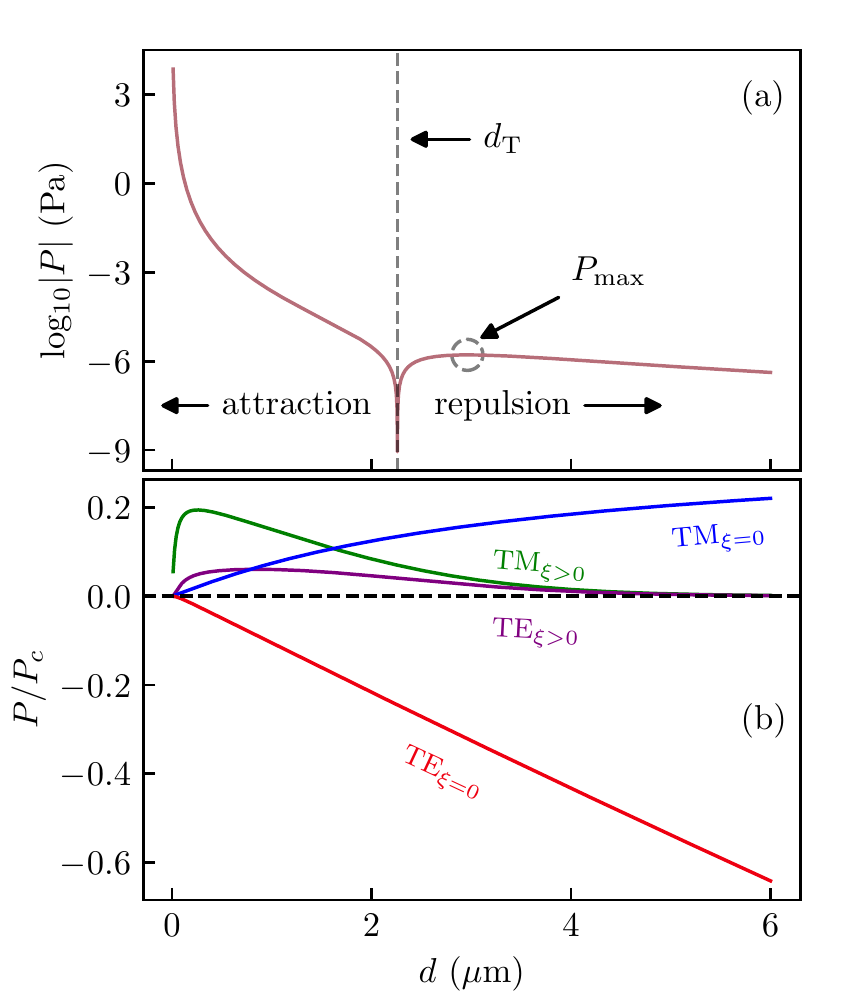}
 	\caption{(a) Magnitude of the pressure between a semi-infinite Au plate and $1 \, \mu$m-thick YIG plate with the transition separation $d_{\text{T}}$ and maximum repulsion $P_{\text{max}}$ features identified. (b) The corresponding mode contributions from the zero-frequency TE and TM modes (TE$_{\xi=0}$, TM$_{\xi=0}$, respectively) and nonzero-frequency TE and TM modes (TE$_{\xi>0}$, TM$_{\xi>0}$, respectively) normalized by $P_{c}$.}
 	\label{fig:basic_Au-YIG}
 \end{figure}
 
For the purposes of identifying typical features in the force behavior, we consider a cavity in vacuum formed by a semi-infinite Au plate and a $1 \, \mu$m-thick YIG plate [$\mu_{\text{YIG}}(0) = 160$~\cite{cagan_fast_1984} and $\epsilon_{\text{YIG}}(0) = 4.02$] at $T = 300 \, \text{K}$. The pressure curve and the separated contributions from the TE and TM modes corresponding to $\xi = 0$ and $\xi > 0$ normalized by the pressure of a perfect conducting cavity in vacuum $P_{c}=-\hbar c \pi^{2}/240 d^{4}$ can be seen in Fig.~\figref{fig:basic_Au-YIG}. 
 
Figure~\figref[a]{fig:basic_Au-YIG} shows the magnitude of the Casimir pressure experienced by the plates as a function of separation. There are two important features here: the vanishing pressure at $d \approx 2.26 \, \mu$m denoting the switch from attraction to repulsion and the largest-magnitude repulsion occurring at $d \approx 2.96 \, \mu$m. We define the separation where this force switching occurs as the transition separation $d_{\text{T}}$ and the largest-magnitude repulsion as $P_{\text{max}}$. These two parameters will be used as metrics for the repulsive behavior in the sections below.
 
 Figure~\figref[b]{fig:basic_Au-YIG} shows the contributions to the pressure from the TE and TM modes at the zero-frequency (TE$_{\xi = 0}$ and TM$_{\xi = 0}$, respectively) and the TE and TM modes for all nonzero frequencies (TE$_{\xi > 0}$ and TM$_{\xi > 0}$, respectively) all normalized by $P_{c}$. The modes at nonzero frequencies control the interaction at small separations, and those at zero-frequency control the interaction at large separations. Interestingly, we note that for these metallic-magnetodielectric systems, the TE$_{\xi=0}$ mode is repulsive.

\section{Relaxed ideality condition on magnetic material} \label{sec:RelaxedIdeality}

The initial cavity considered by Boyer consisted of ideal electric and magnetic plates~\cite{boyer_van_1974}. In real materials, the permittivities and permeabilities are not infinite for all frequencies nor are the two properties mutually exclusive. Some magnetic materials like Ni and Fe are also electrically conductive such that repulsion in such systems may not be present. The question then arises: is there some upper limit on how electrically conductive the magnetic plate can be in order to maintain repulsion? 

\begin{figure}
	\centering
	\includegraphics[width=0.5\textwidth]{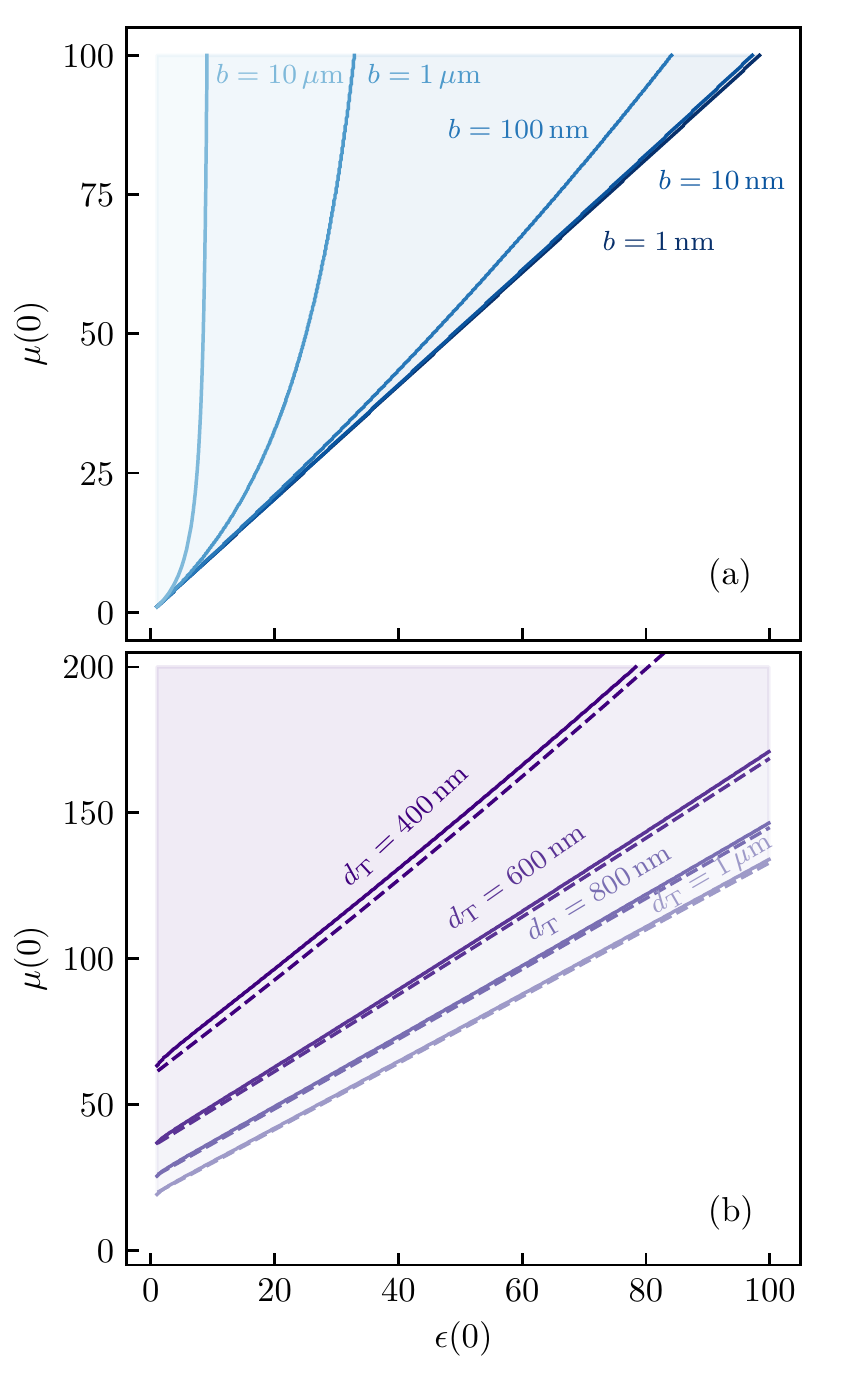}
	\caption{Relations between the zero-frequency dielectric and magnetic response of the magnetic material (a) required to maintain $d_{\text{T}} = 6 \, \mu$m, calculated for varying thicknesses $b$. Above each line are parameter values that result in repulsion; below are those that result in attraction at $d_{\text{T}} = 6 \, \mu$m. (b) Calculations repeated with a single thickness $b = 1 \, \text{nm}$ for varying transition separations $d_{\text{T}}$. The dashed lines correspond to Eq.~\eqref{eqn:analytical_expr}.}
	\label{fig:eps0_vs_mu0}
\end{figure}

To answer this question, for a given value of $\epsilon(0)$, we calculate the value of $\mu(0)$ necessary to keep $d_{\text{T}}$ pinned at $6 \, \mu$m (Fig.~\figref[a]{fig:eps0_vs_mu0}). 
Because the zero-frequency modes dominate the interaction at separations greater than a few microns, this methods allows us to determine the necessary restrictions on the material properties for repulsion in this surface separation regime.

Previous calculations for two infinitely thick plates showed vertically-asymptotic behavior at $\epsilon(0) \approx 8$, putting a relatively short ceiling on how conductive the magnetic material could be~\cite{geyer_thermal_2010}. However, the situation changes as the magnetic material's thickness is reduced to a few nanometers. This asymptotic condition relaxes to a linear relation as the thickness is reduced from 10 $\mu$m to $1 \, \text{nm}$ (Fig.~\figref[a]{fig:eps0_vs_mu0}). Calculations done at $10 \, \text{nm}$ and $1 \, \text{nm}$ have a slope of $\approx1$ in the $\epsilon(0)$-$\mu(0)$ plane, suggesting that the condition for repulsion becomes $\mu(0)\equiv \epsilon(0)$ for very thin films and large separations. 

This simple linear relation could be used as a ``rule of thumb" for generating repulsive Casimir forces at large separations, similar to the inequality between the dielectric functions in fluidic systems~\cite{munday_measured_2009}. As high-resolution experiments are carried out in the sub-micron separation regime, we repeat the calculations for $b = 1 \, \text{nm}$ with $d_{\text{T}} = 1 \, \mu$m, $800 \, \text{nm}$, $600 \, \text{nm}$, and $400 \, \text{nm}$. As shown in Fig.~\figref[b]{fig:eps0_vs_mu0}, the linearity persists, yet the slope increases with a decreasing $d_{\text{T}}$. There is also a constant positive shift in $\mu(0)$ as the repulsive, zero-frequency TE mode has to compensate for the attractive, nonzero-frequency modes that become dominant in the sub-micron separation regime.

To understand the behavior of the numerical curves in Fig.~\figref[b]{fig:eps0_vs_mu0}, we apply several approximations to Lifshitz' formula~\eqref{eqn:pressure} to obtain an analytical expression for $\epsilon(0)$ in terms of $\mu(0)$ which is in agreement with the calculations. For each curve, the total pressure contribution for all non-zero Matsubara frequencies is constant. In the following, we thus apply approximations to the zero-frequency pressure contribution only and keep the contribution of the non-zero Matsubara frequencies exact.

Notice that while the magnetic slab for the data in Fig.~\figref[b]{fig:eps0_vs_mu0} is very thin ($b=1\,$nm), the values of the permittivity $\epsilon(0)$ and permeability $\mu(0)$ are quite high. Thus, there is a competition between physical thickness $b$ and optical thickness for the effective reflection coefficient on the thin magnetic film. The optical thickness is related to the reflection coefficients of the corresponding half-slab $r_{j,0}^\text{R}$. It is high if $\vert r_{j,0}^\text{R}\vert \approx 1$ and low if $r_j^R\approx 0$.

The competition between physical and optical thickness becomes evident when examining the limiting behaviors of small physical thickness and large optical thickness separately.
On the one hand, keeping $r_{j,0}^\text{R}$ fixed and expanding Eq.~\eqref{eqn:eff_fresnel} for small $b$, we obtain $r^{\text{R,eff}}_{j,0} \approx 2kb\,r_{j,0}^{\text{R}}/[1-(r_{j,0}^{\text{R}})^{2}]$. On the other hand, if we keep $b$ fixed and expand Eq.~\eqref{eqn:eff_fresnel} around $r_{j,0}^\text{R}=\pm 1$ where the upper sign corresponds to $j=$TM and the lower to $j=$TE, we find $r^{\text{R,eff}}_{j,0} = \pm 1$. What happens if the physical thickness is small and the optical thickness is large at the same time, as it is the case for the data in Fig.~\figref[b]{fig:eps0_vs_mu0}? The competition between physical and optical thickness can be captured by introducing an effective thickness of the slab by
\begin{equation} \label{eqn:eff_thickness}
    b_{j,\text{eff}}=2b \frac{r_{j,0}^{\text{R}}}{1-(r_{j,0}^{\text{R}})^{2}}\,.
\end{equation}
Notice that for transparent slabs ($\epsilon(0),\mu(0) = 1$) the effective thickness becomes zero as it should. For $\epsilon(0),\mu(0) \gg 1$, the effective thickness becomes $b_{\text{TM},\text{eff}} \approx b \epsilon(0) / 2$ and $b_{\text{TE},\text{eff}} \approx b \mu(0) / 2$ for the two polarizations, respectively. We will make use of this approximation for our final result given in Eq.~\eqref{eqn:analytical_expr}.

If we now keep $b_{j,\text{eff}}$ fixed and expand Eq.~\eqref{eqn:eff_fresnel} around $\vert r_{j,0}^\text{R}\vert =1$ (which is valid for the large values of $\mu(0)$ and $\epsilon(0)$ considered here), we obtain to leading order
\begin{equation}\label{eq:approx_r_j0^Reff}
r_{j,0}^{\text{R,eff}} \approx \frac{kb_{j, \text{eff}}}{kb_{j,\text{eff}}+1}\,.
\end{equation}
Indeed this expression agrees with the limiting behaviors for small physical and large optical thickness discussed above.

As a next step, we approximate the reflection coefficients on the left plate for TE polarization given in~\eqref{eq:r^L}.
As gold has a relatively high plasma frequency, we can expand the reflection coefficient for $\omega_p/c \ll k$ and find
\begin{equation}\label{eq:approx_r^L_TE}
r^\text{L}_\text{TE,0} \approx -1 + 2 \frac{ck}{\omega_p} - 2 \frac{c^2k^2}{\omega_p^2} + \mathcal{O}\left(\frac{c^3k^3}{\omega_p^3}\right)\,.
\end{equation}
The expansion up to second order in $ck/\omega_p$ will be accurate for the separations considered here.

With the approximations~\eqref{eq:approx_r_j0^Reff} and~\eqref{eq:approx_r^L_TE} we can now find an approximation for the pressure contributions at zero frequency.
Introducing the expansion coefficient $\beta_j = b_{j,\text{eff}}/d$, we Taylor expand the pressure contributions up to second order around $\beta_j=0$ and find
\begin{equation}\label{eq:Taylor_approx_pressure}
\begin{aligned}
    P_\text{TE} &\approx \frac{k_B T}{2\pi d^3}\left(\gamma_1 \beta_\text{TE} - \gamma_2\beta_\text{TE}^2 + \mathcal{O}(\beta_\text{TE}^3)\right) \, , \\
    P_\text{TM} &\approx -\frac{k_B T}{2\pi d^3}\left(\frac{3}{8} \beta_\text{TM} - \frac{99}{128}\beta_\text{TM}^2 + \mathcal{O}(\beta_\text{TM}^3)\right) \, ,
\end{aligned}
\end{equation}
with the positive coefficients
\begin{equation}
    \begin{aligned}
        \gamma_1 &= \frac{3}{8}(1 -4\alpha + 10\alpha^2)\, ,\\
        \gamma_2 &= \frac{99}{128}\left(1 - \frac{495}{99}\alpha + 15\alpha^2\right)\, ,
    \end{aligned}
\end{equation}
and $\alpha=c/(d\omega_p)$.

The Taylor expansions \eqref{eq:Taylor_approx_pressure} are accurate, when $\beta_j\ll 1$. Notice that this requirement is actually not quite met for the parameter range considered in Fig.~\figref[b]{fig:eps0_vs_mu0} as the expansion coefficients take values in the range $0\leq \beta_\text{TE}\leq 0.5$ and $0\leq \beta_\text{TE}\leq 0.25$. In fact, expressions \eqref{eq:Taylor_approx_pressure} yield a rather poor approximation to the numerical curves. A better approximation can be found using a Pad\'e approximant. We find
\begin{equation}\label{eq:pade_approx_pressure}
    \begin{aligned}
        P_\text{TE} &\approx \frac{k_B T}{2\pi d^3}\frac{\gamma_1^2 \beta_\text{TE}}{\gamma_2 \beta_\text{TE} + \gamma_1}\,, \\
    P_\text{TM} &\approx -\frac{k_B T}{2\pi d^3}\frac{6 \beta_\text{TM}}{33\beta_\text{TM} + 16}\,.
    \end{aligned}
\end{equation}
The coefficients of the Pad\'e approximants are found by expanding the rational functions around $\beta_j=0$ and comparing the coefficients with the ones from the respective Taylor expansion \eqref{eq:Taylor_approx_pressure}.

Finally, we can leverage the zero-pressure condition that holds true for all calculations included in Fig.~\ref{fig:eps0_vs_mu0},

\begin{equation}
    P^{0}_{\text{TE}} + P^{0}_{\text{TM}} + P^{\xi>0} = 0\,,
	\label{eqn:zero-pressure}
\end{equation}
along with Eqs.~\eqref{eq:pade_approx_pressure} to derive the relation between $\epsilon(0)$ and $\mu(0)$ shown below:
\begin{equation} \label{eqn:analytical_expr}
    \mu(0) \approx \frac{2d}{b} \frac{A\epsilon(0) + B}{C\epsilon(0) + D} \, ,
\end{equation}
with $A=(6+33\delta)\gamma_1 b/(2d),$ $B=16\gamma_1\delta$, $C=[33\gamma_1^2 - (6+33\delta)\gamma_2] b/(2d)$, $D=16(\gamma_1^2 - \gamma_2\delta)$ and $\delta=-2\pi d^3 P^{\xi>0}/(k_B T)$.
The results corresponding to Eq.~\eqref{eqn:analytical_expr} are shown as the dashed lines in Fig.~\figref[b]{fig:eps0_vs_mu0}. It agrees well with the numerical data, validating the linear behavior of the results. The discrepancy between the numerical and analytical results becomes stronger as $d_{\text{T}}$ decreases and is thus most pronounced for the blue curves with $d_{\text{T}}=400\,$nm. This is expected as that curve corresponds to the largest values of $\beta_j$. The accuracy of the analytical results can be improved by calculating a Pad\'e approximant of higher order.


\section{Tuning plate thickness and permeability} \label{sec:TuningMagnetic}

Both the zero-frequency permeability $\mu(0)$ and the plate thickness $b$ are known to affect the repulsion \cite{geyer_thermal_2010, inui_thickness_2011, inui_quantum_2012}. Here we use the transition separation $d_{\text{T}}$ and maximum repulsion $P_{\text{max}}$ (Fig.~\figref[a]{fig:basic_Au-YIG}) as metrics to quantitatively explore their effect and the interplay between these two parameters. In Figs.~\figref[a]{fig:mu0_vs_dtrans,Fmax} and \figref[b]{fig:mu0_vs_dtrans,Fmax}, there is a monotonic decrease in $d_{\text{T}}$ and an increase in $P_{\text{max}}$ as $\mu(0)$ increases. The same relation is seen with a decreasing thickness $b$. 
Our results suggest one should use a material with a large permeability and one that is as thin as possible. 

\begin{figure}
	\centering
	\includegraphics[width=0.5\textwidth]{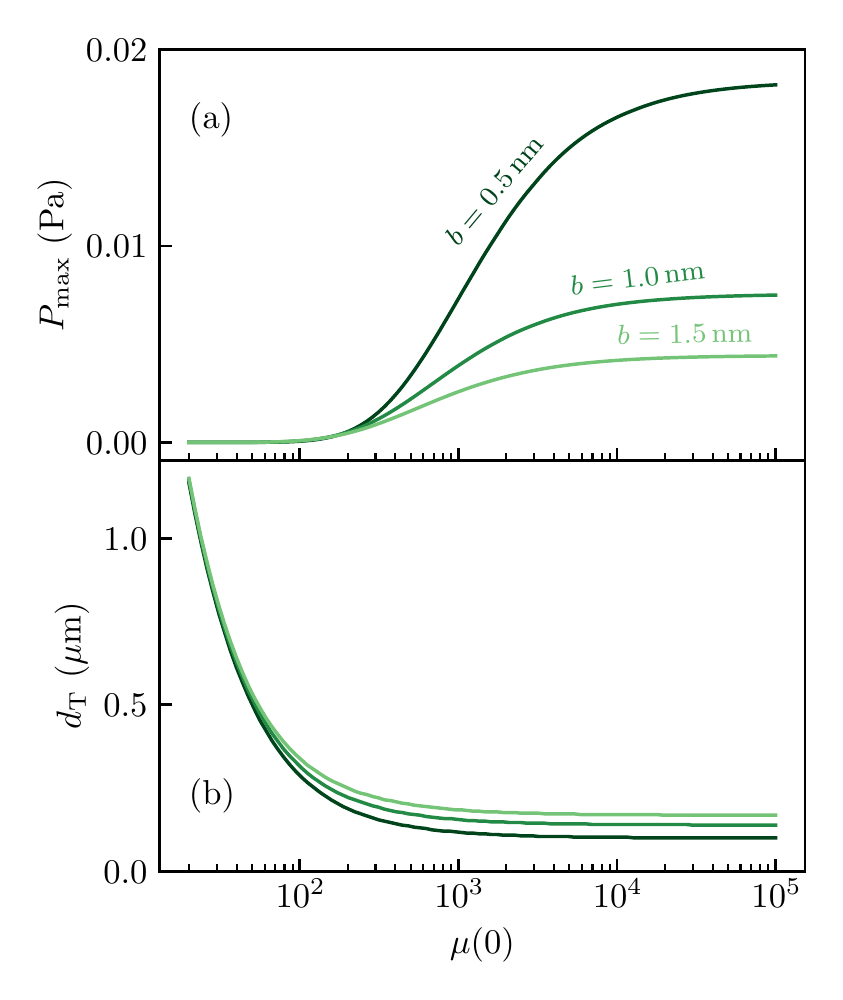}
	\caption{(a) Maximum repulsion $P_{\text{max}}$ and (b) transition separation $d_{\text{T}}$ as a function of the permeability of the magnetic plate $\mu(0)$ for a few plate thicknesses $b$.}
	\label{fig:mu0_vs_dtrans,Fmax}
\end{figure}

\begin{figure}
	\centering
	\includegraphics[width=0.5\textwidth]{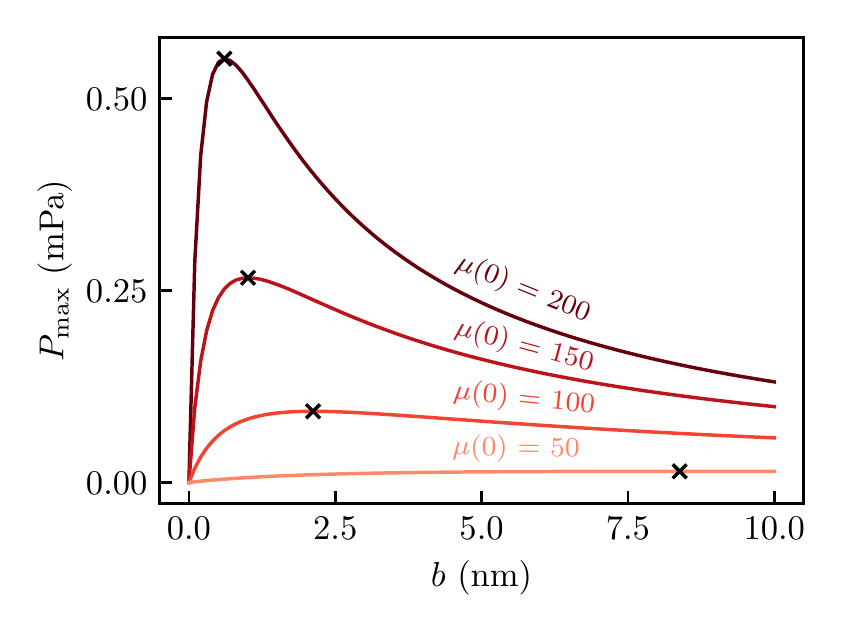}
	\caption{Maximum repulsion $P_{\text{max}}$ calculated as a function of magnetic plate thickness $b$ for a few permeability values $\mu(0)$. The black crosses mark the maxima for each curve.}
	\label{fig:b_vs_Fmax}
\end{figure}

Although $P_{\text{max}}$ appears monotonic with $\mu(0)$, it does not asymptotically approaches $\infty$ as $b \rightarrow 0$. To probe this effect, we keep $\mu(0)$ fixed and vary $b$ (Fig.~\ref{fig:b_vs_Fmax}). $P_{\text{max}}$ increases as $b$ decreases, eventually reaching a peak, beyond which $P_{\text{max}}$ falls towards 0. Further, the peak repulsive pressure shifts to lower thicknesses, and the descent of $P_{\text{max}}$ towards 0 becomes more dramatic for larger values of $\mu(0)$. This result shows that there is such a thing as \textquoteleft too thin,' and the critical thickness where $P_\text{max}$ reaches its peak largely depends on $\mu(0)$. 

As expected, $P_{\text{max}}$ approaches zero as $b \rightarrow 0$, because if the plate thickness goes to zero, the cavity disappears. The initial increase in $P_{\text{max}}$ as $b$ is reduced from bulk thickness can be understood from the different scalings of the TE$_{\xi \rightarrow 0}$ and TM$_{\xi \rightarrow 0}$ pressure contributions with the thickness $b$. The attractive TM$_{\xi \rightarrow 0}$ mode diminishes more quickly than the repulsive TE$_{\xi \rightarrow 0}$ mode with a decreasing $b$, leaving an apparent enhancement to the overall repulsion. This phenomena is described within the context of diamagnetic-ferromagnetic material systems \cite{inui_thickness_2011}, and here, the same argument can be made.

\section{Thermal effects} \label{sec:ThermalEffects}

Beyond the properties of the magnetodielectric material, the system temperature can also be used to enhance repulsion. This phenomena was initially suspected when considering that the temperature can be used to modulate the attraction between two nonmagnetic metals at large separations through the zero-frequency modes \cite{lamoreaux_casimir_2012}. Unlike cavities formed by nonmagnetic metal plates, where an enhanced attraction is seen due to the zero-frequency modes providing attractive contributions, the zero-frequency TE mode in the metallic-magnetodielectric system provides a repulsive contribution and one that dies more slowly with increasing separation than the corresponding attractive TM mode. This connection was confirmed within the context of metallic-metamaterial systems \cite{rosa_casimir_2008}.

Here, we investigate how much we can decrease the transition separation $d_{\text{T}}$ using experimentally-reasonable increases in temperature. The pressure curves in Fig.~\figref[a]{fig:thermal_effects} were calculated using $\mu(0)=20$. Imposing a modest temperature increase from $300 \, \text{K}$ to $310 \, \text{K}$ results in $\Delta d_{\text{T}} = -36.4 \, \text{nm}$. 
In Fig.~\figref[b]{fig:thermal_effects}, pressure curves were again calculated at $300 \, \text{K}$, $305 \, \text{K}$, and $310 \, \text{K}$, but this time with $\mu(0)=160$. With $d_{\text{T}}$ now being much smaller at $\sim240 \, \text{nm}$, the shift in the transition separation driven by the $10 \, \text{K}$ temperature increase is much smaller at $\Delta d_{\text{T}} = -5.6 \, \text{nm}$.

\begin{figure}
	\centering
	\includegraphics[width=0.5\textwidth]{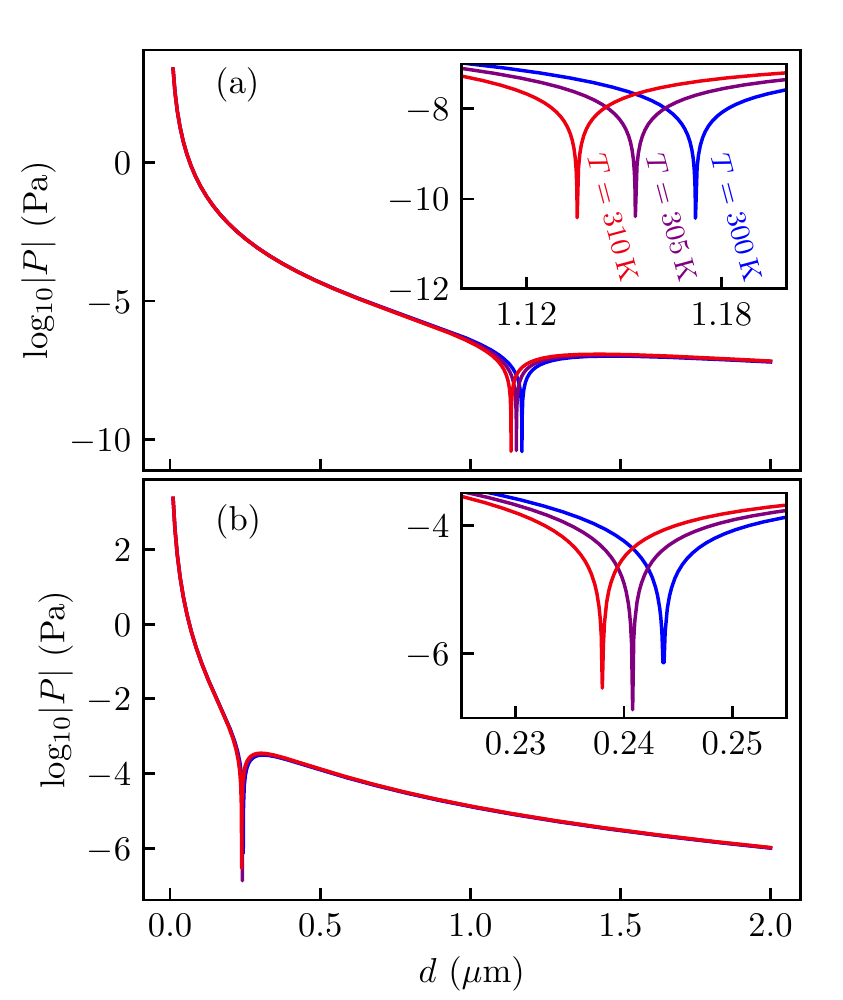}
	\caption{Pressure between Au and a $1 \, \text{nm}$ thick magnetic plate calculated as a function of separation (a) with $\mu(0)=20$ and (b) $\mu(0)=160$. Calculations are performed for temperatures of $300 \, \text{K}$, $305 \, \text{K}$, and $310 \, \text{K}$.}
	\label{fig:thermal_effects}
\end{figure}

From these results, although the gains might be modest, it appears that higher temperatures would be beneficial for pushing the repulsion to lower separations. One must be careful though not to increase the temperature above the Curie point of the magnetodielectric where the thermal excitations will destroy the magnetic ordering. Higher temperatures will also increase the thermal measurement noise~\cite{garrett_sensitivity_2019}.

\section{Conclusions} \label{sec:Conclusions}

We have shown how three parameters related to the magnetic plate [$\epsilon(0)$, $\mu(0)$, and $b$] and the system temperature $T$ can be leveraged to enhance and modulate the repulsion in metallic-magnetodielectric plate systems. Our results concerning the ideality of the magnetic material are consistent with previous work~\cite{geyer_thermal_2010}; however, we show that when the thickness $b$ is decreased to only a few nm, the required relation between $\mu(0)$ and $\epsilon(0)$ for generating repulsion at large separations is $\mu(0) = \epsilon(0)$ and stays linear well-into the sub-micron separation regime. Further, a large magnetic permeability (preferably $\mu(0) > 1000$) and a small thickness ($b \sim 1 \, \text{nm}$) can drive repulsion on a mPa-scale within a few-hundred nanometer separation regime, although there is a lower limit on how thin the material can be before the pressure decreases. Lastly, high temperatures (but below the Curie point) are optimal as they can boost the repulsive zero-frequency TE mode contribution.

From this investigation, we identify magnetic van der Waals materials \cite{burch_magnetism_2018, gong_two-dimensional_2019} as excellent candidates for these systems, given their nm-scale and even sub-nm-scale thicknesses. These materials could potentially be used to generate repulsive Casimir forces in vacuum (or air) using current techniques \cite{de_man_casimir_2010, garrett_sensitivity_2019} and even levitated above a Au sample, similar to the experiment seen in Ref.~\cite{zhao_stable_2019} using a fluid. Beyond the potential measurement of repulsive forces in air, subsequent measurements could provide additional experimental evidence for the appropriateness of plasma or Drude models when discussing Casimir forces, see e.g., Refs.~\cite{banishev_demonstration_2013, bimonte_isoelectronic_2016, bimonte_measurement_2021}.

\section*{Acknowledgments}

The authors acknowledge financial support from the Defense Advanced Research Projects Agency (DARPA) QUEST Projects Contract No. HR00112090084. CS acknowledges support by the National Science Foundation Graduate Research Fellowship Program under Grant No. 2036201. Any opinions, findings, and conclusions or recommendations expressed in this material are those of the author(s) and do not necessarily reflect the views of the National Science Foundation or DARPA.

\bibliography{references}

\end{document}